\documentclass{article}

\usepackage{amsmath}
\usepackage{natbib}
\usepackage[title]{appendix}
\usepackage[margin=1.0in]{geometry}
\usepackage{times}

\numberwithin{equation}{section}

\title{\bf On the dissociation between potential vorticity conservation and symmetries}

\author{Martin Charron and Ayrton Zadra \\
	Recherche en pr\'evision num\'erique atmosph\'erique \\
	Environnement et Changement climatique Canada, Dorval, Qc, Canada}

\date{\today}

\begin{document}

\maketitle

\begin{abstract}
Using a four-dimensional manifestly covariant formalism suitable for classical fluid dynamics, it is shown that the conservation of potential vorticity is not associated with any symmetry of the equations of motion but is instead a trivial conservation law of the second kind. The demonstration is provided in arbitrary coordinates and therefore applies to comoving (or label) coordinates. Since this is at odds with previous studies, which claimed that potential vorticity conservation is associated with a symmetry under particle-relabeling, a detailed discussion on relabeling transformations is also presented.
\end{abstract}

\section{Introduction} \label{intro}
The equations of motion of physical systems may always be written in a covariant form because the laws governing their evolution do not depend on the choice of coordinates. To render covariance explicit, governing equations of classical fluid mechanics may be written in terms of tensor components suitable for any curvilinear, non-inertial coordinate system in which time intervals are absolute \citep{Charron14a}. This ensures that the underlying theory is relativistic \citep{Charron15b}, obviously not in the traditional sense of describing fluids with velocities comparable to that of light but in the literal sense of ``obeying a principle of relativity''---in this case, the principle of Newtonian relativity.

In such formulations, conservation laws take the same form in all coordinate systems:
\begin{align}
\frac{\partial}{\partial x^\mu}\left(\sqrt{g} A^\mu\right)=\frac{\partial}{\partial t}\left(\sqrt{g} A^0\right)+\frac{\partial}{\partial x^i}\left(\sqrt{g} A^i\right)=0, \label{cl}
\end{align}
where $A^\mu$ is a conserved 4-current with $A^0$ a scalar charge density and $A^i$ its flux density. The quantity $g$ is the determinant of the covariant metric tensor $g_{\mu\nu}$ of the coordinate system. Repeated Greek indices are summed from $0$ to $3$, and repeated Latin indices from $1$ to $3$. The coordinate $x^0=t$ is the time, and the $x^i$'s are any curvilinear spatial coordinates. An invariant spatial volume element is written $dx^1dx^2dx^3\sqrt{g} \equiv d^3x\sqrt{g}$.

The equations of motion of an inviscid classical fluid in an external gravitational potential $\Phi$ under adiabatic conditions may be expressed as
\begin{align}
\Lambda^\mu&=0, \label{mmge} \\
\Lambda_{(\beta)}&=0, \label{ege}
\end{align}
where the 4-vector $\Lambda^\mu$ and the scalar $\Lambda_{(\beta)}$ are
\begin{align}
\Lambda^\mu& \equiv \left(\rho u^\mu u^\nu + h^{\mu\nu}p\right)_{:\nu} + \rho h^{\mu\nu} \Phi_{,\nu}, \label{mmeq} \\
\Lambda_{(\beta)}& \equiv -\rho u^\mu s_{,\mu} \label{ls}
\end{align}
(the parentheses around the symbol $\beta$ indicate that it is not a space-time index---see \citet{Zadra15} where these equations are derived from a least action principle). The symbol $\rho$ represents the fluid density, $u^\mu \equiv dx^\mu/dt$ the 4-velocity field with $u^0=1$, $h^{\mu\nu}\equiv g^{\mu\nu}-g^{0\mu}g^{0\nu}/g^{00}$ a symmetric contravariant tensor for purely spatial distances with $h^{0\mu}=h^{\mu 0}=0$, $p$ the pressure, and $s$ the fluid specific entropy. The absolute nature of time intervals in Newtonian mechanics imposes a constraint on the space-time metric tensor: the contravariant component $g^{00}$ must be a non-zero constant (taken here as unity). A comma followed by an index indicates an ordinary partial derivative: $(\cdot)_{,\mu}\equiv \partial (\cdot)/\partial x^\mu$. A material derivative is written $d(\cdot)/dt\equiv u^\mu (\cdot)_{,\mu}$. A colon followed by an index indicates a covariant derivative. For instance, given a 4-vector $A^\mu$, its covariant derivative ${A^\mu}_{:\nu}$ is
\begin{align}
{A^\mu}_{:\nu} \equiv {A^\mu}_{,\nu} + \Gamma^{\mu}_{\alpha\nu}A^\alpha.
\end{align}
The terms
\begin{align}
\Gamma^{\mu}_{\alpha\nu} \equiv \frac{1}{2} g^{\mu\beta}(g_{\alpha\beta,\nu}+g_{\beta\nu,\alpha}-g_{\nu\alpha,\beta})
\end{align}
are Christoffel symbols of the second kind. Because time intervals are absolute in Newtonian mechanics, $\Gamma^0_{\alpha\nu}=0$. Moreover, the absolute nature of time intervals implies that a contravariant tensor of rank $n$ becomes a tensor of rank $n-m$ ($0 \le m \le n$) when $m$ of its indices are set to zero. For instance, if $B^{\mu\nu}$ is a contravariant second-rank tensor, $B^{0\nu}$ is a contravariant first-rank tensor (a contravariant 4-vector), and $B^{00}$ is a zeroth-rank tensor (a scalar). Notice that, in general, this rule does not apply to covariant tensors.

The equation $\Lambda^0=(\rho u^\nu)_{:\nu}=0$ represents mass continuity, $\Lambda^i=0$ the three momentum equations, and $\Lambda_{(\beta)}=0$ the material conservation of specific entropy. Because $\sqrt{g}\,\Gamma^\mu_{\nu\mu}=(\sqrt{g})_{,\nu}$, a conservation equation such as \eqref{cl} is equivalent to $(\sqrt{g}A^\mu)_{,\mu}=\sqrt{g}{A^\mu}_{:\mu}=0$. The formalism used in this paper is described in more details in \citet{Charron14a}.

A relation is said to hold {\it on-shell} when the equations of motion \eqref{mmge} and \eqref{ege} are used to obtain the relation. It is said to hold {\it off-shell} when $\Lambda^\mu$ and $\Lambda_{(\beta)}$ are not necessarily assumed to vanish.

Previous work by \citet{Newcomb67,Bretherton70,Ripa81,Salmon82,Salmon88,Salmon98,Salmon13,Muller95,Padhye96} and others associated the conservation of potential vorticity with a particular symmetry transformation of the equations of motion. In the following, it is argued that this association is unjustified because potential vorticity conservation (Ertel's theorem) is a trivial law of the second kind, following the terminology of \citet{Olver93}. Such trivial laws are obtained independently of the equations of motion, and therefore are not associated with symmetries of these equations. The statement that potential vorticity conservation is a trivial law may perhaps surprise fluid dynamicists. Here, ``trivial'' obviously does not mean dynamically uninteresting or useless. As will be seen below, it means in mathematical terms that the conserved current associated with potential vorticity is given by the divergence of an antisymmetric tensor, and that it is not associated with a symmetry of the dynamics.

In section \ref{comoving}, a specific sub-class of admissible coordinate systems called comoving coordinates is described. Comoving coordinates are characterized by a dynamic mesh that follows exactly the moving fluid elements. This sub-class of coordinates is introduced because it is used by authors interested in the particle-relabeling transformation. In section \ref{revisit}, the use of Noether's first theorem under particle-relabeling is revisited. The triviality of potential vorticity conservation in arbitrary coordinates is then demonstrated in section \ref{trivial}. This establishes that potential vorticity conservation cannot be related to a symmetry of the equations of motion. Starting from arbitrary coordinates, it is straightforward to show the triviality of potential vorticity conservation in comoving coordinates. This is also done in section \ref{trivial}. Conclusions are drawn in section \ref{conclu}.

\section{Comoving coordinates} \label{comoving}
The tensor formalism introduced in the preceding section applies to any coordinate system admissible in classical fluid mechanics. In this section, a particular sub-class of reference frames called comoving coordinates is described. Such coordinate systems are ``attached'' to the fluid elements and move with the fluid. Therefore, the three spatial coordinates associated with a given fluid element, often called its labels, do not change as time evolves. One may start from the tensor formalism in arbitrary coordinates described in section \ref{intro} and find specific relations that apply in comoving coordinates only. As is well known, such coordinates are useful to express a Lagrangian for fluid dynamics in a traditional form, i.e.\ kinetic minus potential energy \citep{Herivel55}.

Comoving coordinates will be referred to as $\hat{x}^\mu$, with $\hat{x}^0=\tau=t$ the time and $\hat{x}^i$ the labels. In the following, comoving coordinates and quantities expressed in comoving coordinates will always be indicated with hatted variables, except scalars---say $f$, for which $\hat{f}=f$ at all space-time points. Because the comoving coordinate mesh follows the fluid elements, the 4-velocity $\hat{u}^\mu \equiv d\hat{x}^\mu/dt=(1,0,0,0)$. The mass continuity equation and the material conservation of specific entropy, respectively \eqref{mmge} with $\mu=0$ and \eqref{ege}, take the form
\begin{align}
\sqrt{\hat{g}} \Lambda^0 &= \frac{\partial}{\partial \hat{x}^0} \left(\sqrt{\hat{g}}\rho\hat{u}^0\right) + \frac{\partial}{\partial \hat{x}^i} \left(\sqrt{\hat{g}}\rho \hat{u}^i\right)=\frac{\partial}{\partial \tau} \left(\sqrt{\hat{g}}\rho\right)=0, \label{contcom} \\
\Lambda_{(\beta)} &= -\rho\left( \hat{u}^0\frac{\partial s}{\partial \hat{x}^0}+\hat{u}^i\frac{\partial s}{\partial \hat{x}^i}\right)=-\rho \frac{\partial s}{\partial \tau}=0, \label{lbhat}
\end{align}
where $\hat{g}$ is the determinant of the covariant metric tensor $\hat{g}_{\mu\nu}$ in coordinates $\hat{x}^\mu$. The product $\sqrt{\hat{g}}\rho$ as well as specific entropy $s$ are therefore time-independent on-shell in comoving coordinates. It may be shown from \eqref{mmeq} that one form of the on-shell equations governing the covariant 4-velocity is written
\begin{align}
\frac{\hat{\Lambda}_j-\hat{u}_j\Lambda^0}{\rho}=\frac{\partial \hat{u}_j}{\partial \tau}+\frac{1}{\rho}\frac{\partial p}{\partial \hat{x}^j}+\frac{\partial}{\partial \hat{x}^j}\left( \Phi-\frac{1}{2}\hat{g}_{00}\right)=0,
\end{align}
where
\begin{align}
\frac{\hat{\Lambda}_j-\hat{u}_j\Lambda^0}{\rho}=\frac{\partial x^\nu}{\partial \hat{x}^j}g_{\mu\nu}\left(\frac{\Lambda^\mu-u^\mu\Lambda^0}{\rho}\right)=\frac{\partial x^i}{\partial \hat{x}^j}\left(\frac{\Lambda_i-u_i\Lambda^0}{\rho}\right)
\end{align}
with
\begin{align}
\frac{\Lambda_i-u_i\Lambda^0}{\rho}=\frac{du_i}{dt}-\frac{1}{2}g_{\mu\nu,i}u^\mu u^\nu +\frac{1}{\rho}p_{,i} + \Phi_{,i}.
\end{align}

\section{Revisiting Noether's first theorem under particle-relabeling} \label{revisit}
In a particle-like formulation of fluid dynamics, potential vorticity conservation has in the past been associated with the particle-relabeling transformation by \citet{Newcomb67,Bretherton70,Ripa81,Salmon82,Salmon88,Salmon98,Salmon13,Muller95,Padhye96} and others. Noether's first theorem is often invoked to justify the association between the particle-relabeling transformation and potential vorticity conservation. This association is here challenged.

In this section, a demonstration of Noether's first theorem under a particle-relabeling transformation is presented. Special attention is given to one crucial step in the demonstration; a step where identically vanishing terms must be identified or else one will be misled to wrong conclusions.

In a particle-like formulation, the fluid particles may be labeled with any admissible curvilinear coordinate system $(\hat{x}^1,\hat{x}^2,\hat{x}^3)$ at a given time $\hat{x}^0=\tau$. The mass of a given fluid element $dm$ is provided by $d\hat{x}^1d\hat{x}^2d\hat{x}^3\sqrt{\hat{g}} \rho =d^3\hat{x}\sqrt{\hat{g}} \rho$. Imposing the constancy in time of $dm$ ensures the {\it a priori} conservation of total mass and is equivalent to $\partial (\sqrt{\hat{g}} \rho)/\partial \tau=0$. In addition, the material conservation of specific entropy must also be assumed {\it a priori}, therefore $s \equiv s(\hat{x}^1,\hat{x}^2,\hat{x}^3)$ and $\partial s/\partial \tau=0$. In other words, in a particle-like formulation of Hamilton's least action principle it is always assumed that $\Lambda^0$ and $\Lambda_{(\beta)}$ vanish {\it a priori}\footnote{This assumption is unnecessary when using Clebsch potentials as dynamical fields, see e.g.\ \citet{Zadra15}.}. Therefore, in this section the expression ``off-shell'' concerns the momentum equations only.

Consider Noether's first theorem for the particle-like formulation in comoving coordinates. The action functional is written
\begin{align}
{\cal S}=\int d\tau \, d^3\hat{x} \, \sqrt{\hat{g}} \rho (K-I-\Phi), \label{action}
\end{align}
with $K$ the scalar kinetic energy density per unit mass as calculated in an inertial frame, and $I=I(\rho,s)$ the internal energy density per unit mass of the fluid. This action functional leads to the three momentum equations of motion but does not lead to mass and entropy conservation---which are assumed {\it a priori}. In this formulation, the dynamical fields are the spatial coordinates $x^i$ in an arbitrary frame (but not comoving), and therefore
\begin{align}
K=\frac{1}{2} \hat{g}_{\mu\nu}\hat{u}^\mu\hat{u}^\nu=\frac{1}{2} g_{\mu\nu}u^\mu u^\nu=\frac{1}{2} g_{\mu\nu}\frac{\partial x^\mu}{\partial \tau}\frac{\partial x^\nu}{\partial \tau} \nonumber
\end{align}
since
\begin{align}
u^\mu=\frac{\partial x^\mu}{\partial \hat{x}^\nu}\hat{u}^\nu=\frac{\partial x^\mu}{\partial \hat{x}^0}\hat{u}^0+\frac{\partial x^\mu}{\partial \hat{x}^i}\hat{u}^i=\frac{\partial x^\mu}{\partial \tau}.
\end{align}

A particle-relabeling transformation will be interpreted as a passive coordinate transformation from $(\hat{x}^1,\hat{x}^2,\hat{x}^3)$ to $(\hat{x}'^1,\hat{x}'^2,\hat{x}'^3)$ within the sub-class of comoving coordinates, where
\begin{align}
\hat{x}'^0 &= \tau' = \hat{x}^0 = \tau = t, \\
\hat{x}'^i &= \hat{x}^i + \hat{\epsilon}^i
\end{align}
are also comoving coordinates, and with $\hat{\epsilon}^0=0$ and $\hat{\epsilon}^i$ infinitesimally small. Because both $\hat{x}'^\mu$ and $\hat{x}^\mu$ are chosen to be comoving coordinates with $d\hat{x}'^i/dt=d\hat{x}^i/dt=0$, it means that $\hat{\epsilon}^i=\hat{\epsilon}^i(\hat{x}^1,\hat{x}^2,\hat{x}^3)$ must be independent of $\tau$. A passive variation of the action functional ${\cal S}$ under such particle-relabeling is written
\begin{align}
\tilde{\delta} {\cal S} = \int d\tau d^3\hat{x} \left[ \tilde{\delta}(\sqrt{\hat{g}} \rho)(K-I-\Phi) + \sqrt{\hat{g}} \rho (\tilde{\delta}K - \tilde{\delta}I -\tilde{\delta}\Phi) \right]. \label{labelds}
\end{align}
Under an infinitesimal particle-relabeling, the dynamical fields (the arbitrary, non-comoving coordinates) $x^i(\tau,\hat{x}^1,\hat{x}^2,$ $\hat{x}^3)$ become $\tilde{x}^i(\tau',\hat{x}'^1,\hat{x}'^2,\hat{x}'^3)$. The symbol $\tilde{x}^i$ indicates a different functional form induced by the particle-relabeling, however the values of the actual arbitrary coordinates at a given point do not change under a particle-relabeling: $x^i(\tau,\hat{x}^1,\hat{x}^2,\hat{x}^3)=\tilde{x}^i(\tau',\hat{x}'^1,\hat{x}'^2,\hat{x}'^3)=\tilde{x}^i(\tau,\hat{x}^1,\hat{x}^2,\hat{x}^3)+(\partial x^i/\partial \hat{x}^j) \, \hat{\epsilon}^j$ to first order. The variation induced on the dynamical fields $x^i$ by a particle-relabeling interpreted as a passive transformation is therefore given by
\begin{align}
\tilde{\delta}x^i \equiv \tilde{x}^i(\tau,\hat{x}^1,\hat{x}^2,\hat{x}^3)-x^i(\tau,\hat{x}^1,\hat{x}^2,\hat{x}^3)=-\frac{\partial x^i}{\partial \hat{x}^j}\hat{\epsilon}^j.
\end{align}

Passive transformations in comoving coordinates leave $d^3\hat{x}$ unchanged. Since the mass of a fluid element is conserved {\it a priori}, the passive transformation of a mass element is $\tilde{\delta}(d^3\hat{x} \sqrt{\hat{g}}\rho)=\tilde{\delta}(\sqrt{\hat{g}}\rho)d^3\hat{x}=0$, implying that
\begin{align}
\tilde{\delta}(\sqrt{\hat{g}}\rho)=0. \label{labeldm}
\end{align}

Under a particle-relabeling transformation, since $K-I-\Phi$ is a scalar, the passive variation of $\sqrt{\hat{g}} \rho (K-I-\Phi)$ is
\begin{align}
\tilde{\delta} \left( \sqrt{\hat{g}} \rho (K-I-\Phi) \right)&= \sqrt{\hat{g}} \rho \tilde{\delta}(K-I-\Phi) = -\sqrt{\hat{g}} \rho \frac{\partial (K-I-\Phi)}{\partial \hat{x}^i} \hat{\epsilon}^i, \nonumber \\
&=-\frac{\partial}{\partial \hat{x}^i} \left(  \sqrt{\hat{g}} \rho \hat{\epsilon}^i (K-I-\Phi) \right)+(K-I-\Phi)\frac{\partial}{\partial \hat{x}^i} \left(  \sqrt{\hat{g}} \rho \hat{\epsilon}^i \right). \label{varl1}
\end{align}
The passive variation of $\sqrt{\hat{g}} \rho (K-I-\Phi)$ may also be expressed as
\begin{align}
\tilde{\delta} \left( \sqrt{\hat{g}} \rho (K-I-\Phi) \right)&= \sqrt{\hat{g}}\rho \left[ \frac{1}{2}u^\mu u^\nu \tilde{\delta} g_{\mu\nu} + u_j \tilde{\delta} u^j -\tilde{\delta}I-\tilde{\delta}\Phi  \right], \nonumber \\
&=\sqrt{\hat{g}}\rho \hat{\epsilon}^i \left[ -\frac{1}{2}u^\mu u^\nu \frac{\partial g_{\mu\nu}}{\partial \hat{x}^i} -u_j \frac{\partial}{\partial \tau} \left( \frac{\partial x^j}{\partial \hat{x}^i}\right)+ \frac{p}{\rho^2} \frac{\partial \rho}{\partial \hat{x}^i} + \frac{\partial \Phi}{\partial \hat{x}^i}\right] \nonumber \\
&\quad + \sqrt{\hat{g}}\rho \hat{\epsilon}^i T\frac{\partial s}{\partial \hat{x}^i}, \nonumber \\
&=\sqrt{\hat{g}}\rho \hat{\epsilon}^i \frac{\partial x^j}{\partial \hat{x}^i} \left[ \frac{\partial u_j}{\partial \tau}-\frac{1}{2}u^\mu u^\nu \frac{\partial g_{\mu\nu}}{\partial x^j} + \frac{1}{\rho}\frac{\partial p}{\partial x^j} + \frac{\partial \Phi}{\partial x^j}\right] \nonumber \\
& \quad - \sqrt{\hat{g}}\rho \hat{\epsilon}^i \frac{\partial}{\partial \hat{x}^i}\left( \frac{p}{\rho} \right) + \sqrt{\hat{g}}\rho \frac{\partial x^j}{\partial \hat{x}^i} u_j \frac{\partial \hat{\epsilon}^i}{\partial \tau} + \sqrt{\hat{g}}\rho \hat{\epsilon}^i T\frac{\partial s}{\partial \hat{x}^i}\nonumber \\
& \quad -\frac{\partial}{\partial \tau} \left( \sqrt{\hat{g}}\rho \hat{\epsilon}^i \frac{\partial x^j}{\partial \hat{x}^i} u_j \right), \nonumber \\
&=-\sqrt{\hat{g}} \Lambda_j \tilde{\delta} x^j + \frac{\partial}{\partial \tau} \left( \sqrt{\hat{g}} \rho u_j \tilde{\delta} x^j \right) + \sqrt{\hat{g}}\rho \frac{\partial x^j}{\partial \hat{x}^i} u_j \frac{\partial \hat{\epsilon}^i}{\partial \tau} \nonumber \\
& \quad -\frac{\partial}{\partial \hat{x}^i} \left( \sqrt{\hat{g}} \hat{\epsilon}^i p  \right) + \frac{p}{\rho} \frac{\partial}{\partial \hat{x}^i} \left( \sqrt{\hat{g}}\rho \hat{\epsilon}^i \right)+ \sqrt{\hat{g}}\rho \hat{\epsilon}^i T\frac{\partial s}{\partial \hat{x}^i}, \label{varl2}
\end{align}
where $T$ is temperature. Equating \eqref{varl1} with \eqref{varl2}, one obtains Noether's first theorem under passive transformations in comoving coordinates:
\begin{align}
\frac{\partial}{\partial \tau} &\left( \sqrt{\hat{g}}\rho u_j \tilde{\delta} x^j \right) + \frac{\partial}{\partial \hat{x}^i} \left( \sqrt{\hat{g}} \rho \hat{\epsilon}^i \left[ K-\Phi-I-\frac{p}{\rho} \right]  \right) \nonumber \\
&= \sqrt{\hat{g}} \Lambda_j \tilde{\delta} x^j + \left(K-\Phi-I-\frac{p}{\rho}  \right) \frac{\partial}{\partial \hat{x}^i} \left( \sqrt{\hat{g}} \rho \hat{\epsilon}^i \right)-\sqrt{\hat{g}} \rho \frac{\partial x^j}{\partial \hat{x}^i} u_j \frac{\partial \hat{\epsilon}^i}{\partial \tau} \nonumber \\
&\quad \; - T \sqrt{\hat{g}} \rho \hat{\epsilon}^i \frac{\partial s}{\partial \hat{x}^i}. \label{nrl}
\end{align}
The three conditions
\begin{align}
\tilde{\delta}(\sqrt{\hat{g}}\rho)=-\frac{\partial}{\partial \hat{x}^i} \left( \sqrt{\hat{g}} \rho \hat{\epsilon}^i \right)&=0, \label{lc1} \\
\tilde{\delta}s=-\frac{\partial s}{\partial \hat{x}^i}\hat{\epsilon}^i&=0, \label{lc2} \\
\frac{\partial \hat{\epsilon}^i}{\partial \tau}&=0 \label{lc3}
\end{align}
established by \citet{Padhye96} must be satisfied to obtain a conservation equation on-shell. The expressions \eqref{lc1} and \eqref{lc2} are nothing but constraints on, respectively, mass\footnote{The first equality sign in \eqref{lc1} follows from the relations $\tilde\delta(\sqrt{\hat{g}})=-\partial(\sqrt{\hat{g}}\hat{\epsilon}^i)/\partial\hat{x}^i$ and $\tilde\delta\rho=-\hat{\epsilon}^i \partial\rho/\partial\hat{x}^i$.} and entropy conservation under a passive comoving coordinate transformation. These two constraints impose that $\Lambda^0$ and $\Lambda_{(\beta)}$ continue to vanish {\it a priori}. They are unrelated to symmetries of the dynamics obtained from minimizing the action functional \eqref{action}, which leads to the momentum equations only---not mass and entropy conservation.

Notice that \eqref{lc1} and \eqref{lc2} may be expressed as covariant equations in arbitrary coordinates:
\begin{align}
\frac{1}{\sqrt{\hat{g}}} \tilde{\delta} (\sqrt{\hat{g}}\rho)=-\frac{1}{\sqrt{\hat{g}}}\frac{\partial}{\partial \hat{x}^i} \left( \sqrt{\hat{g}} \rho \hat{\epsilon}^i \right)=-\frac{1}{\sqrt{\hat{g}}}\frac{\partial}{\partial \hat{x}^\mu} \left( \sqrt{\hat{g}} \rho \hat{\epsilon}^\mu \right)=-(\rho\epsilon^\mu)_{:\mu}&=0, \\
\tilde{\delta}s=-\frac{\partial s}{\partial \hat{x}^i}\hat{\epsilon}^i=-\frac{\partial s}{\partial \hat{x}^\mu}\hat{\epsilon}^\mu=-s_{,\mu}\epsilon^\mu&=0,
\end{align}
given that $\hat{\epsilon}^\mu={{\hat{x}}^\mu}_{\, \; ,\nu}\epsilon^\nu$ and $\hat{\epsilon}^0=\epsilon^0=0$. However, the condition \eqref{lc3} is not covariant under arbitrary coordinate transforma\-tions---its covariant form would be $\partial \hat{\epsilon}^i/\partial \tau + \hat{\Gamma}^i_{0j}\hat{\epsilon}^j=0$. Therefore, \eqref{lc3} is satisfied in comoving coordinates only. This lack of covariance implies that \eqref{lc3} has no specific meaning in arbitrary coordinates. As is the case for \eqref{lc1} and \eqref{lc2}, the constraint \eqref{lc3} is unrelated to a symmetry of the dynamics and only imposes that the transformed coordinates remain comoving with the fluid. The choice
\begin{align}
\hat{\epsilon}^i = \frac{\varepsilon^{0ijk}}{\sqrt{\hat{g}}\rho} \frac{\partial s}{\partial \hat{x}^j} \frac{\partial \zeta}{\partial \hat{x}^k}, \label{rls}
\end{align}
with $\zeta=\zeta(\hat{x}^1,\hat{x}^2,\hat{x}^3)$ an arbitrary infinitesimal passive tracer, leads to an acceptable transformation in these coordinates since it satisfies the three constraints \eqref{lc1}--\eqref{lc3}. The term $\varepsilon^{\mu\nu\alpha\beta}$ is the Levi-Civita symbol, and $(\sqrt{\hat{g}})^{-1}\varepsilon^{\mu\nu\alpha\beta}$ is a contravariant fourth-rank tensor with vanishing covariant derivative.

The term $\sqrt{\hat{g}}\rho u_j \tilde{\delta} x^j$ appearing in \eqref{nrl} may be rewritten as
\begin{align}
\sqrt{\hat{g}}\rho u_j \tilde{\delta} x^j=\sqrt{\hat{g}}\rho q\zeta+\frac{\partial}{\partial \hat{x}^j}\left( \varepsilon^{0ijk} \hat{u}_i \frac{\partial s}{\partial \hat{x}^k}\zeta\right)
\end{align}
from the definition of potential vorticity $q$:
\begin{align}
q \equiv \frac{\varepsilon^{0ijk}}{\sqrt{\hat{g}}\rho}\frac{\partial \hat{u}_j}{\partial \hat{x}^i} \frac{\partial s}{\partial \hat{x}^k}.
\end{align}
Define
\begin{align}
\hat{A}^{ij}&\equiv \frac{\varepsilon^{0ijk}}{\sqrt{\hat{g}}} \left(K-\Phi-I-\frac{p}{\rho} \right)\frac{\partial s}{\partial \hat{x}^k}=-\hat{A}^{ji}, \label{aij} \\
\hat{B}^{i}&\equiv \frac{\varepsilon^{0ijk}}{\sqrt{\hat{g}}} \hat{u}_j \frac{\partial s}{\partial \hat{x}^k}, \label{gbi} \\
\hat{b}^i&\equiv - \frac{\varepsilon^{0ijk}}{\sqrt{\hat{g}}} \frac{\hat{\Lambda}_j}{\rho} \frac{\partial s}{\partial \hat{x}^k}. \label{bi}
\end{align}
Off-shell and once the three constraints are explicitly satisfied, \eqref{nrl} reduces to
\begin{align}
\frac{\partial}{\partial \tau} \left( \sqrt{\hat{g}}\rho q\zeta \right) + \frac{\partial}{\partial \hat{x}^j} \left[ \sqrt{\hat{g}}\frac{\partial \zeta}{\partial \hat{x}^i} \hat{A}^{ij} - \frac{\partial}{\partial \tau}\left(\sqrt{\hat{g}}\hat{B}^j\zeta\right) \right] = \sqrt{\hat{g}}\frac{\partial \zeta}{\partial \hat{x}^i}\hat{b}^i, \label{rlnoether}
\end{align}
from the transformation \eqref{rls}, and from \eqref{contcom} and \eqref{lbhat}. The relation \eqref{rlnoether} is a conservation law in comoving coordinates on-shell (i.e.\ it has the form of a continuity equation in comoving coordinates when $\hat{b}^i=0$). However due to the presence of $\zeta$, \eqref{rlnoether} on-shell {\it is not Ertel's theorem}.

One may then follow the line of thought presented in \citet{Padhye96}, and manipulate \eqref{rlnoether} to rewrite it as
\begin{align}
\zeta \left[ \frac{\partial}{\partial \tau}\left(\sqrt{\hat{g}}\rho q\right) +\frac{\partial}{\partial \hat{x}^i} \left(\sqrt{\hat{g}}\hat{b}^i\right)\right] = \frac{\partial}{\partial \hat{x}^i} \left[ \sqrt{\hat{g}}\zeta \hat{b}^i + \sqrt{\hat{g}}\frac{\partial \zeta}{\partial \hat{x}^j}\hat{A}^{ij} + \zeta\frac{\partial}{\partial \tau} \left(\sqrt{\hat{g}}\hat{B}^i\right) \right] \label{pm96}
\end{align}
after using the condition $\partial \zeta / \partial \tau=0$. \citet{Padhye96}, working on-shell (i.e.\ $\hat{b}^i=0$), integrate this equation over the labels, use the divergence theorem with suitable boundary conditions, and conclude from the du Bois-Reymond lemma that potential vorticity is materially conserved {\it because} $\zeta$ is arbitrary.

We however arrive at a fundamentally different conclusion. It will be shown in sub-section \ref{incomoving} that both the left-hand side and right-hand side of \eqref{pm96} vanish {\it identically} when mass and entropy are assumed to be conserved {\it a priori}. This implies that the arbitrariness of $\zeta$, the du Bois-Reymond lemma, and therefore Noether's first theorem are irrelevant to the fact that the material derivative of potential vorticity vanishes on-shell. Given that \eqref{pm96} is nothing but the algebraic identity $0=0$ when mass and entropy are conserved {\it a priori}, this relation does not represent a non-trivial conservation law that requires the equations of motion for momentum and specific symmetry conditions.

\section{Triviality of potential vorticity conservation} \label{trivial}
In this section, it will be demonstrated that the equation describing potential vorticity conservation, which is written in the form of a continuity equation as \eqref{cl}, is an algebraic identity off-shell. This implies that it exists independently of any symmetry.

\subsection{Definition of a trivial conservation law of the second kind}
First, consider a generic antisymmetric tensor $F^{\mu\nu}=-F^{\nu\mu}$. Define a 4-vector $c^\mu$ as the covariant divergence of this antisymmetric tensor:
\begin{align}
c^\mu\equiv {F^{\mu\nu}}_{:\nu}. \nonumber
\end{align}
By virtue of \eqref{i3}, the covariant divergence of $c^\mu$ identically vanishes:
\begin{align}
{c^\mu}_{:\mu}={F^{\mu\nu}}_{:\nu:\mu}=0. \label{secondkind}
\end{align}
Following the terminology of \citet[][p.\ 264-265]{Olver93}, \eqref{secondkind} is a trivial conservation law of the second kind. Such trivial conservation laws are characterized by currents solely written in terms of the divergence of an antisymmetric tensor. They are algebraic identities obtained off-shell---i.e.\ the equations of motion \eqref{mmge} and \eqref{ege} are not required to establish such trivial conservation laws---and therefore they exist independently of symmetries of the equations of motion.

Antisymmetric quantities such as $F^{\mu\nu}$ are sometimes referred to as ``superpotentials''. Conserved currents may always be defined up to the divergence of a superpotential.

\subsection{In arbitrary coordinates}
In tensor notation, potential vorticity $q$ is defined as
\begin{align}
q \equiv \frac{\omega^\mu s_{,\mu}}{\rho},
\end{align}
where
\begin{align}
\omega^\mu &= \frac{\varepsilon^{0\mu \nu \alpha}}{\sqrt{g}} u_{\alpha:\nu}=\frac{\varepsilon^{0\mu \nu \alpha}}{\sqrt{g}} u_{\alpha,\nu}, \\ \label{omeg}
u_\alpha &= g_{\alpha\mu}u^\mu.
\end{align}

Here, the fields defining $q$ are considered off-shell, i.e.\ they are not assumed to be governed by \eqref{mmge} and \eqref{ege}. The trivial nature of potential vorticity conservation may be demonstrated in arbitrary coordinates. It is shown that
\begin{align}
c^\mu=\rho u^\mu q + \frac{\omega^\mu \Lambda_{(\beta)}}{\rho} - \frac{\varepsilon^{0\mu \nu \alpha}}{\sqrt{g}} \frac{(\Lambda_\nu-u_\nu \Lambda^0)s_{,\alpha}}{\rho} = {F^{\mu\nu}}_{:\nu}, \label{curqobs}
\end{align}
where
\begin{align}
F^{\mu\nu}  &= u^\mu B^\nu - u^\nu B^\mu + \frac{\varepsilon^{0\mu \nu \alpha}}{\sqrt{g}} \left[ \left(\frac{1}{2}u^\sigma u_\sigma-\Phi-I-\frac{p}{\rho} \right)s_{,\alpha} + u_\alpha \frac{\Lambda_{(\beta)}}{\rho} \right], \label{gbm} \\
B^\mu         &= \frac{\varepsilon^{0\mu \nu \alpha}}{\sqrt{g}} u_\nu s_{,\alpha} \label{bbb}
\end{align}
(see Appendix \ref{trivobs}). The fact that $F^{\mu\nu}$ is antisymmetric ensures that $c^\mu$ in \eqref{curqobs} is a trivially conserved 4-current of the second kind. The charge density $c^0$ associated with that 4-current is $\rho q$ off-shell because $\varepsilon^{00\nu\alpha}=0$. The conservation law
\begin{align}
\frac{\partial}{\partial t}(\sqrt{g} \rho q) + \frac{\partial}{\partial x^i} \left( \sqrt{g} \left[ \rho u^i q + \frac{\omega^i \Lambda_{(\beta)}}{\rho} - \frac{\varepsilon^{0ijk}}{\sqrt{g}} \frac{(\Lambda_j-u_j \Lambda^0)s_{,k}}{\rho}\right]\right)=0, \label{troglo}
\end{align}
which follows from taking the covariant divergence of \eqref{curqobs}, reduces to $(\sqrt{g}\rho u^\mu q)_{,\mu}=\sqrt{g}\rho u^\mu q_{,\mu}=\sqrt{g}\rho\, dq/dt=0$ on-shell---i.e.\ after using the equations of motion $\Lambda_j=g_{j\nu}\Lambda^{\nu}=0$, $\Lambda^0=0$ and $\Lambda_{(\beta)}=0$. The conservation law \eqref{troglo} is an algebraic identity and is demonstrated without assuming that the equations of motion are satisfied and therefore without assuming any symmetry of the equations of motion.

The triviality of potential vorticity conservation has been suggested in previous studies. For instance, \citet{Muller95} demonstrated that the evolution of the potential vorticity charge density $\rho q$ is governed by a conservation equation which is a mathematical identity, although he still associated it with particle-relabelling. \citet{Rosenhaus16} also analyzed the triviality of potential vorticity conservation in the context of incompressible flows, but did not discuss particle-relabeling transformations.

\subsection{In comoving coordinates} \label{incomoving}
The antisymmetric tensor provided by \eqref{gbm} may be expressed in comoving coordinate systems with $\hat{u}^i=0$ as
\begin{align}
\hat{F}^{00}&=0, \label{hatF1} \\
\hat{F}^{0i}&=\frac{\varepsilon^{0ijk}}{\sqrt{\hat{g}}}\hat{u}_j\frac{\partial s}{\partial \hat{x}^k}=-\hat{F}^{i0}, \label{hatF2} \\
\hat{F}^{ij}&=\frac{\varepsilon^{0ijk}}{\sqrt{\hat{g}}} \left[ \left(K-\Phi-I-\frac{p}{\rho} \right)\frac{\partial s}{\partial \hat{x}^k} + \hat{u}_k \frac{\Lambda_{(\beta)}}{\rho} \right]=-\hat{F}^{ji}. \label{hatF3}
\end{align}
The 4-current \eqref{curqobs} may also be expressed in comoving coordinates as
\begin{align}
\hat{c}^0&=\frac{1}{\sqrt{\hat{g}}}\frac{\partial}{\partial \hat{x}^\nu}\left(\sqrt{\hat{g}}\hat{F}^{0\nu}\right)=\rho q, \label{hatc1} \\
\hat{c}^i&=\frac{1}{\sqrt{\hat{g}}}\frac{\partial}{\partial \hat{x}^\nu}\left(\sqrt{\hat{g}}\hat{F}^{i\nu}\right)=\frac{\hat{\omega}^i \Lambda_{(\beta)}}{\rho} - \frac{\varepsilon^{0ijk}}{\sqrt{\hat{g}}} \frac{(\hat{\Lambda}_j-\hat{u}_j \Lambda^0)}{\rho} \frac{\partial s}{\partial \hat{x}^k}. \label{hatc2}
\end{align}

The trivial conservation law of the second kind \eqref{troglo} for potential vorticity density is valid in arbitrary coordinates. In particular, it may be expressed in comoving coordinates with $\hat{u}^i=0$:
\begin{align}
\frac{\partial}{\partial \tau}(\sqrt{\hat{g}} \rho q) + \frac{\partial}{\partial \hat{x}^i} \left( \sqrt{\hat{g}} \left[ \frac{\hat{\omega}^i \Lambda_{(\beta)}}{\rho} - \frac{\varepsilon^{0ijk}}{\sqrt{\hat{g}}} \frac{(\hat{\Lambda}_j-\hat{u}_j \Lambda^0)}{\rho}\frac{\partial s}{\partial \hat{x}^k}\right]\right)=0. \label{tricom}
\end{align}
On-shell, the terms $\Lambda_{(\beta)}$, $\hat{\Lambda}_j$, and $\Lambda^0$ all vanish and this trivial conservation law becomes $\partial (\sqrt{\hat{g}} \rho q)/\partial \tau=0$. From \eqref{contcom}, this reduces to $\partial q/\partial \tau=0$, i.e.\ the material conservation of potential vorticity.

In comoving coordinates as in any other admissible coordinate system, the material conservation of potential vorticity is simply an on-shell version of a trivial law of the second kind. Therefore, it cannot be associated with a particle-relabeling transformation or any symmetry of the equations of motion because trivial conservation laws of the second kind exists independently of these governing equations. Particle-relabeling is merely a coordinate transformation within the sub-class of comoving coordinates whose associated constraints \eqref{lc1}--\eqref{lc3} are unrelated to the dynamics obtained from minimizing \eqref{action} and its symmetries. Because the equations of motion are covariant, a particle-relabeling transformation does leave their form intact but does not imply a dynamically relevant conservation equation, in particular for potential vorticity.

Consider now $\hat{F}^{ij}$ under the assumptions that $\Lambda^0$ and $\Lambda_{(\beta)}$ vanish {\it a priori}, as was required in section \ref{revisit}. It is nothing but $\hat{A}^{ij}$ provided by \eqref{aij}. Under the same assumptions, $\hat{F}^{0i}$ becomes $\hat{B}^i$ from \eqref{gbi}, and $\hat{c}^i$ becomes $\hat{b}^i$ from \eqref{bi}. The left-hand side of \eqref{pm96} is therefore written
\begin{align}
\zeta \left[ \frac{\partial}{\partial \tau}\left(\sqrt{\hat{g}}\rho q\right) +\frac{\partial}{\partial \hat{x}^i} \left(\sqrt{\hat{g}}\hat{b}^i\right)\right]=\zeta \left[ \frac{\partial}{\partial \hat{x}^0} \left( \sqrt{\hat{g}} \hat{c}^0 \right) + \frac{\partial}{\partial \hat{x}^i}\left( \sqrt{\hat{g}} \hat{c}^i \right) \right]=\zeta \frac{\partial^2}{\partial \hat{x}^\mu\partial\hat{x}^\nu} \left( \sqrt{\hat{g}} \hat{F}^{\mu\nu} \right)=0.
\end{align}
It vanishes identically due to the commutativity of ordinary derivatives and the antisymmetry of $\hat{F}^{\mu\nu}$. The right-hand side of \eqref{pm96} also vanishes identically:
\begin{align}
\frac{\partial}{\partial \hat{x}^i} \left[ \sqrt{\hat{g}}\zeta \hat{b}^i + \sqrt{\hat{g}}\frac{\partial \zeta}{\partial \hat{x}^j}\hat{A}^{ij} + \zeta \frac{\partial}{\partial \tau} \left(\sqrt{\hat{g}}\hat{B}^i\right) \right]&=\frac{\partial}{\partial \hat{x}^i} \left[ \sqrt{\hat{g}}\zeta \hat{c}^i + \sqrt{\hat{g}}\frac{\partial \zeta}{\partial \hat{x}^j}\hat{F}^{ij} + \zeta \frac{\partial}{\partial \hat{x}^0} \left(\sqrt{\hat{g}}\hat{F}^{0i}\right) \right], \nonumber \\
&=\frac{\partial \zeta}{\partial \hat{x}^i}\left[ \sqrt{\hat{g}} \hat{c}^i + \frac{\partial}{\partial \hat{x}^j}\left( \sqrt{\hat{g}} \hat{F}^{ji}\right)+ \frac{\partial}{\partial \hat{x}^0} \left(\sqrt{\hat{g}}\hat{F}^{0i}\right) \right] \nonumber \\
&\quad +\zeta \left[ \frac{\partial}{\partial\hat{x}^i}\left( \sqrt{\hat{g}}\hat{c}^i\right) +\frac{\partial^2}{\partial\hat{x}^0\partial\hat{x}^i} \left(\sqrt{\hat{g}}\hat{F}^{0i}\right) \right], \nonumber \\
&=\frac{\partial \zeta}{\partial \hat{x}^i}\left[ \sqrt{\hat{g}} \hat{c}^i - \frac{\partial}{\partial \hat{x}^\nu}\left( \sqrt{\hat{g}} \hat{F}^{i\nu}\right) \right] \nonumber \\
&\quad +\zeta \left[ \frac{\partial}{\partial\hat{x}^i}\left( \sqrt{\hat{g}}\hat{c}^i\right) +\frac{\partial}{\partial\hat{x}^0} \left(\sqrt{\hat{g}}\hat{c}^0\right) \right], \nonumber \\
&=\frac{\partial \zeta}{\partial \hat{x}^i}\left[ \sqrt{\hat{g}} \hat{c}^i - \sqrt{\hat{g}} \hat{c}^i \right]+\zeta \frac{\partial^2}{\partial \hat{x}^\mu\partial\hat{x}^\nu}\left( \sqrt{\hat{g}} \hat{F}^{\mu\nu}\right), \nonumber \\
&=0.
\end{align}
This implies that, given {\it a priori} conservation of mass and entropy, \eqref{pm96} is always true, whether the constraint \eqref{lc3} associated with particle-relabeling is satisfied or not. In a similar fashion, it may be verified that \eqref{rlnoether} identically reduces to
\begin{align}
\sqrt{\hat{g}}\hat{B}^i\frac{\partial}{\partial \hat{x}^i}\left( \frac{\partial\zeta}{\partial\tau} \right)=0
\end{align}
when mass and entropy are conserved {\it a priori}, implying that Noether's first theorem under particle-relabeling is unrelated to potential vorticity conservation but only associates particle-relabeling to the material conservation of $\zeta$.

The application of Noether's first theorem simply leads to the consistency check of a constraint, given that potential vorticity conservation is a trivial law of the second kind obtained without assuming any symmetry. Noether's first theorem under a particle-relabeling is a circular statement on the material conservation of $\zeta$, not a statement on the material conservation of $q$. If one ignores the fact that \eqref{troglo} is a trivial conservation law, one will wrongly associate particle-relabeling with potential vorticity conservation.

It has previously been mentioned that the supposed symmetry associated with potential vorticity conservation is invisible in an Eulerian formulation but exists in a label formulation \citep[see for example][]{Shepherd15}. The demonstration above with a manifestly covariant formulation---which applies equally to Eulerian and comoving (or label) coordinates---clearly shows that this is not the case. There is no hidden or apparent symmetry associated with potential vorticity conservation.

\section{Summary and conclusions}\label{conclu}
A trivial conservation law of the second kind is an off-shell identity obtained independently of the equations of motion and {\it a fortiori} of any assumed symmetry of these equations. A conservation law cannot be trivial of the second kind, and at the same time exist on-shell as a consequence of a continuous symmetry of the equations of motion. In this paper, it was demonstrated that potential vorticity conservation is a trivial law of the second kind in arbitrary coordinates and is therefore dissociated from symmetries. Consequently, the association of potential vorticity conservation with the particle-relabeling transformation made by several authors is unfounded. In this paper, a typical mistake made by authors associating potential vorticity conservation with particle-relabeling has been pointed out.

Particle-relabeling is naturally defined in terms of comoving coordinate transformations. One of its associated constraints is not covariant under arbitrary coordinate transformations. Continuous symmetries or constraints that are apparent only in a given sub-class of coordinate systems but are broken in arbitrary coordinates---i.e.\ resulting from non-covariant conditions---cannot give rise to dynamically relevant conservation laws. This is because the covariance of the equations of motion implies that if a non-trivial symmetry of these equations exists, it must exist independently of the choice of coordinates---inertial, non-inertial, Eulerian, comoving, etc.

Particle-relabeling is parameterized by an arbitrary passive tracer $\zeta$. From Noether's first theorem under particle-relabeling, the associated conserved charge density is $\rho q \zeta$. Given that $\rho q$ was shown to be the charge density of a trivial conservation law (i.e.\ obtained independently of particle-relabeling or any symmetry transformation), it follows that the conservation law associated with particle-relabeling is simply equivalent to the material conservation of $\zeta$, which does nothing but confirm the prior assumption that $\zeta$ was a passive tracer.

\section*{Acknowledgements} The authors are grateful to Christopher Subich and St\'ephane Gaudreault for providing comments that led to an improved manuscript.

\begin{appendices}
\section{Two tensor identities}
\begin{enumerate}
\item Consider an antisymmetric tensor $F^{\mu\nu}=-F^{\nu\mu}$. It will be shown that the scalar ${F^{\mu\nu}}_{:\nu:\mu}$ vanishes in a Riemannian space. The term $\sqrt{g}{F^{\mu\nu}}_{:\nu}$ is written
\begin{align}
\sqrt{g}{F^{\mu\nu}}_{:\nu}&=\sqrt{g}{F^{\mu\nu}}_{,\nu}+\sqrt{g}\,\Gamma^\mu_{\alpha\nu}F^{\alpha\nu}+\sqrt{g}\,\Gamma^\nu_{\alpha\nu}F^{\mu\alpha}, \nonumber \\
&=\sqrt{g}{F^{\mu\nu}}_{,\nu}+(\sqrt{g})_{,\nu}F^{\mu\nu}, \nonumber \\
&=(\sqrt{g}F^{\mu\nu})_{,\nu}. \label{ii3}
\end{align}
Moreover,
\begin{align}
{F^{\mu\nu}}_{:\nu:\mu} &= (\sqrt{g})^{-1}(\sqrt{g}{F^{\mu\nu}}_{:\nu})_{,\mu}, \nonumber \\
&=(\sqrt{g})^{-1}(\sqrt{g}F^{\mu\nu})_{,\nu,\mu}, \nonumber \\
&=0 \label{i3}
\end{align}
from \eqref{ii3}, the antisymmetry of $F^{\mu\nu}$, and the commutativity of ordinary derivatives.
\item One may verify the identity
\begin{align}
\varepsilon^{\alpha\mu\nu\sigma}{A^\beta}_{\alpha} +  \varepsilon^{\beta\alpha\nu\sigma}{A^\mu}_{\alpha} + \varepsilon^{\beta\mu\alpha\sigma}{A^\nu}_{\alpha} + \varepsilon^{\beta\mu\nu\alpha}{A^\sigma}_{\alpha} =\varepsilon^{\beta\mu\nu\sigma}{A^\alpha}_{\alpha},
\end{align}
contract it with $\delta^0_\beta$, and replace ${A^\mu}_{\alpha}$ by ${u^\mu}_{:\alpha}$ to obtain
\begin{align}
\varepsilon^{0\alpha\nu\sigma}{u^\mu}_{:\alpha} + \varepsilon^{0\mu\alpha\sigma}{u^\nu}_{:\alpha} + \varepsilon^{0\mu\nu\alpha}{u^\sigma}_{:\alpha}=\varepsilon^{0\mu\nu\sigma}{u^\alpha}_{:\alpha}. \label{i2}
\end{align}
This holds because ${u^0}_{:\alpha}=0$ (recall that $u^0=1$ and $\Gamma^0_{\mu\nu}=0$). This is a tensor identity when multiplied by $(\sqrt{g})^{-1}$.
\end{enumerate}

\section{Potential vorticity conservation as a trivial law} \label{trivobs}
In this Appendix, potential vorticity conservation expressed in arbitrary coordinates is shown to be a trivial law of the second kind. From the definition of potential vorticity,
\begin{align}
\rho u^\alpha q&=\frac{\varepsilon^{0\mu\nu\sigma}}{\sqrt{g}}u_{\nu:\mu}s_{,\sigma} u^\alpha, \nonumber \\
&= \left( \frac{\varepsilon^{0\mu\nu\sigma}}{\sqrt{g}} u_\nu s_{,\sigma} u^\alpha  \right)_{:\mu} - \frac{\varepsilon^{0\mu\nu\sigma}}{\sqrt{g}} u_\nu s_{,\sigma} {u^\alpha}_{:\mu}, \nonumber \\
&= \left( u^\alpha B^\mu \right)_{:\mu} + \frac{1}{\sqrt{g}} u_\nu s_{,\sigma} \left( \varepsilon^{0\alpha\mu\sigma} {u^\nu}_{:\mu} + \varepsilon^{0\alpha\nu\mu} {u^\sigma}_{:\mu} - \varepsilon^{0\alpha\nu\sigma} {u^\mu}_{:\mu}\right), \nonumber \\
&= \left( u^\alpha B^\mu - u^\mu B^\alpha + \frac{\varepsilon^{0\alpha\mu\sigma}}{\sqrt{g}} \left[ s_{,\sigma} K + u_\sigma \frac{\Lambda_{(\beta)}}{\rho}  \right] \right)_{:\mu} -\omega^\alpha \frac{\Lambda_{(\beta)}}{\rho} \nonumber \\
&\qquad + u^\mu {B^\alpha}_{:\mu}-\frac{\varepsilon^{0\alpha\nu\mu}}{\sqrt{g}} u_\nu u^\sigma s_{,\sigma:\mu} \nonumber
\end{align}
after using \eqref{i2}, \eqref{ls}, \eqref{omeg}, \eqref{bbb} and $K=u_\nu u^\nu /2$. From
\begin{align}
u^\mu {B^\alpha}_{:\mu}-\frac{\varepsilon^{0\alpha\nu\mu}}{\sqrt{g}} u_\nu u^\sigma s_{,\sigma:\mu} &= \frac{\varepsilon^{0\alpha\mu\sigma}}{\sqrt{g}} u^\beta u_{\mu:\beta} s_{,\sigma}, \nonumber \\
&= \frac{\varepsilon^{0\alpha\mu\sigma}}{\sqrt{g}} \frac{(\Lambda_\mu-u_\mu \Lambda^0)}{\rho}s_{,\sigma} -\left( \frac{\varepsilon^{0\alpha\mu\sigma}}{\sqrt{g}} s_{,\sigma} \Phi \right)_{:\mu} \nonumber \\
&\qquad -\frac{\varepsilon^{0\alpha\mu\sigma}}{\sqrt{g}} \frac{p_{,\mu}}{\rho} s_{,\sigma}, \nonumber \\
&= \frac{\varepsilon^{0\alpha\mu\sigma}}{\sqrt{g}} \frac{(\Lambda_\mu-u_\mu \Lambda^0)}{\rho}s_{,\sigma} -\left( \frac{\varepsilon^{0\alpha\mu\sigma}}{\sqrt{g}} s_{,\sigma} \Phi \right)_{:\mu} \nonumber \\
&\qquad -\left( \frac{\varepsilon^{0\alpha\mu\sigma}}{\sqrt{g}} (I+p/\rho) s_{,\sigma} \right)_{:\mu},
\end{align}
one finally gets
\begin{align}
\rho u^\alpha q&=\left( u^\alpha B^\mu - u^\mu B^\alpha + \frac{\varepsilon^{0\alpha\mu\sigma}}{\sqrt{g}} \left[ s_{,\sigma} (K-\Phi-I-p/\rho) + u_\sigma \frac{\Lambda_{(\beta)}}{\rho}  \right] \right)_{:\mu} \nonumber \\
&\qquad -\omega^\alpha \frac{\Lambda_{(\beta)}}{\rho} + \frac{\varepsilon^{0\alpha\mu\sigma}}{\sqrt{g}} \frac{(\Lambda_\mu-u_\mu \Lambda^0)s_{,\sigma}}{\rho},
\end{align}
which proves \eqref{curqobs}.

\end{appendices}

\bibliographystyle{ametsoc2014}
\bibliography{refer}

\end{document}